\documentclass[amstex,epsf,amssymb,usenatbib,letters]{mn2e}

\usepackage{epsfig}
\usepackage{graphicx}
\usepackage{tabularx}
\usepackage{amsmath}
\usepackage{amssymb}
\usepackage{amstext}
\usepackage{amsfonts}

\usepackage{soul}
\usepackage{xcolor}
\setulcolor{cyan}   
\setstcolor{red}    
\sethlcolor{yellow} 

\title{Density distributions of outflow driven turbulence}

\author[A. Moraghan, Jongsoo Kim \& Suk-Jin Yoon]
{Anthony Moraghan $^{1,2}$,
Jongsoo Kim $^{2}$,
\& Suk-Jin Yoon $^{1}$\thanks{E-mail: sjyoon@galaxy.yonsei.ac.kr},\\
$^1$Center for Galaxy Evolution Research and Department of Astronomy, Yonsei University, Seoul 120-749, Republic of Korea\\
$^2$Korea Astronomy and Space Science Institute, 61-1, Hwaam-dong, Yuseong-gu, Daejeon 305-348, Republic of Korea\\ 
}

\date{Accepted 2013 March 26
      Received 2013 March 25 ;
      in original form 2013 Februrary 7}

\pagerange{\pageref{firstpage}--\pageref{lastpage}}
\pubyear{2013}

\begin{document}

\maketitle

\label{firstpage}

\begin{abstract}
Protostellar jets and outflows are signatures of star formation and 
promising mechanisms for driving supersonic turbulence in molecular clouds.
We quantify outflow-driven turbulence through three-dimensional numerical simulations
using an isothermal version of the robust total variation diminishing code.
We drive turbulence in real-space using a simplified spherical outflow model, analyse the data 
through density probability distribution functions (PDF), and investigate the Core Formation Rate per free-fall time (CFR$_{ff}$).
The real-space turbulence driving method produces a negatively skewed density PDF possessing an enhanced tail on the low-density side.
It deviates from the log-normal distributions typically obtained from Fourier-space turbulence driving 
at low densities, but can provide a good fit at high-densities, particularly in terms of mass weighted rather than volume weighted 
density PDF.
Due to this fact, we suggest that the CFR$_{ff}$ determined from a Fourier-driven turbulence model could be comparable to that 
of our particular real-space driving model, which has a ratio of solenoidal to compressional components from the 
resulting turbulence velocity fields of $\sim$0.6.
\end{abstract}


\section{Introduction}              

Protostellar jets and outflows are important by-products of the early stages of star formation where a collapsing 
protostellar system expels excess accretion material in order to lose angular momentum. 
This material takes the form of bipolar protostellar jets, which are collimated and accelerated along 
the rotation axis of the system. The jets entrain ambient cloud material 
to form protostellar outflows which propagate and deposit energy and momentum up to several parsecs into the 
molecular cloud\,\citep[e.g.][]{2004A&A...415..189M}.
Protostellar jets and outflows are thus promising candidates for turbulence generation in molecular clouds 
as they drive it internally and may enable star formation to be a self-regulating process.

Although simulations of protostellar outflows clearly show a transfer of momentum from the jet to the ambient 
medium \citep[e.g.][]{2008MNRAS.386.2091M},
previous studies have suggested they may not be capable of sustaining turbulence in molecular clouds.
\citet{2007ApJ...668.1028B} simulated single jets and analysed the data using velocity 
probability distribution functions (PDF) to find that
supersonic fluctuations decay quickly and do not spread far from the jet.
They found {\it subsonic} non-compressional modes occupy the disturbed volume instead.
Later,\,\citet{2009ApJ...692..816C} showed that by taking a more realistic non-uniform ambient 
environment consisting of pre-existing turbulent motions or `fossil cavities', then the outflow 
morphology can be significantly modified leading to a more efficient transfer of energy to the ambient medium.
Recently, \citet{2009ApJ...695.1376C,2010ApJ...722..145C} built upon the previous work with simulations of multiple 
interacting outflows in a global parsec scale volume analysed via power-spectra. 
Their simulations suggest that outflows can sustain turbulence through their interactions.

On the observational side, 
in a recent study by \citet{2012MNRAS.425.1380I} of the Serpens and Aquila regions, the authours estimated 
the outflows they observed would be unable to support the turbulence of the surrounding molecular clouds.
In contrast, in a study of the Perseus cloud complex, \citet{2010ApJ...715.1170A} found that a significant 
fraction of the turbulence can be sustained by outflows. 
Similarly, \citet{2012ApJ...746...25N} who surveyed the L1641-N cluster suggest that 
outflow feedback can influence the dynamical evolution of the clump.

Since the work of \citet{1994ApJ...423..681V} it is widely accepted that for isothermal supersonic turbulence,
the density PDF is log-normal with a standard deviation, $\sigma$, proportional to the Mach number $\mathcal{M}$. 
The relationship is now usually defined through the relation 
$\sigma^2 = \mathrm{ln} \left(1+b^2 \mathcal{M}^2\right)$, as first formulated by \citet{1997MNRAS.288..145P} with $b = 0.5$. 
Other groups find similar results but with different $b$ values (See \citet{2008ApJ...688L..79F} and references therein).
Such log-normal fits readily appear, not only in many numerical simulations of turbulence, but also through analytical analysis,
and observational data of molecular clouds where observers measure column-densities of tracer molecules\,\citep[e.g.][]{2010MNRAS.406.1350F}.

Many recent works on numerical simulations of turbulence now drive it in Fourier-space 
through a combination of a solenoidal and compressional based forcing 
mechanism\,\citep[e.g.][]{2008ApJ...688L..79F,2010A&A...512A..81F,2012MNRAS.423.2680M}. 
\citet{2008ApJ...688L..79F} 
find that the addition of compressive forcing makes the standard-deviation of the density PDF 
about 3 times wider than that of solenoidal driving alone, and 
also confirm the relationship between standard-deviation and Mach number.
However, little research has been performed relating to the density PDF from simulations of turbulence 
driven in real-space.

In this work we investigate the concept of outflow driven turbulence numerically using a modified version of the
isothermal total variation diminishing ({\sc tvd}) code. 
Whereas most previous turbulence simulations drive the turbulence in Fourier-space, we use a simple model
considering protostellar outflows to drive the turbulence real-space.
We characterise the resulting turbulence in terms of density PDF and core formation rate per free fall time (CFR$_{ff}$).
We find that driving the turbulence in real-space via outflows, leads to density PDFs with negatively skewed 
log-normal fits possessing enhanced tails at the low-density side.
This deviates from the log-normal distributions commonly found in isothermal driven turbulence simulations.
However, due to good agreement on the high-density side of the density PDF, particularly in terms of mass, 
the results suggest that estimates of the CFR$_{ff}$ based purely 
on log-normal fits, or from Fourier driven turbulence, could still provide comparable results to those 
of real-space driven turbulence.

The layout of our paper is as follows. 
We describe the basis of our model for generating turbulence and the computational code used in this study in \S~\ref{method}. 
The numerical results and density PDFs are presented and discussed in \S~\ref{results}.
We summarise and conclude our findings in \S~\ref{conclusion}.

\section{Numerical Methods}            
\label{method}

Observations of many active molecular clouds, such as the Serpens and Aquila region, 
have revealed many complex dynamical processes 
including starless cores, young stellar objects, outflows, and turbulence \citep[e.g.][]{2012MNRAS.425.1380I}.
Modelling a realistic molecular cloud would be very comprehensive, requiring an external 
medium to sufficiently bind it, as well as magnetic fields, molecular physics, self-gravity, and a 
star formation accretion-ejection mechanism.

As a first step, we study a simplified case as we are trying to understand the properties of 
turbulence driven by, and associated with, outflows only. 
We therefore inject a series of spherical outflows without self-gravity or magnetic field effects into 
the computational domain.
As such, spherical outflows were considered by \citet{2006ApJ...640L.187L} and \citet{2007ApJ...659.1394M}.
Although \citet{2007ApJ...659.1394M} claim collimated outflows are more efficient than
spherical outflows at driving turbulence as they can propagate further from their source thus
driving turbulence on larger scales which decay slower, in this work our purpose is to start 
with a simple model.

Our turbulence forcing model proceeds as follows; A 3D computational domain is initialised with a uniform gaseous medium.
Different points within the domain are randomly selected. At each randomly selected point, the mean 
density, $\bar{\rho}$, of a volume defined by the outflow radius, $R$, is determined. 
If the mean density of this volume is larger than the global mean, $\rho_0$, and no point within the 
volume is below a lower-threshold value 0.001$\rho_0$, then it is chosen to be the location of an outflow.

The momentum, $P$, which an outflow should introduce is defined as follows,
\begin{equation}
 P = \iiint\limits_{<R} \rho (\textbf{\textit{r}} ) v_j (R) \left( \frac{r}{R} \right)^q \mathrm{d} V
\label{P}
\end{equation}
where $\rho (\textbf{\textit{r}})$ is the density within the outflow volume depending on distance vector $\textbf{\textit{r}}$ from the 
outflow center, and $v_j(R)$ is the outflow velocity at its outermost boundary, $R$. 
The $q$ parameter, set to $1.0$, attempts to mimic the observed `Hubble-law' type velocity
observed in protostellar outflows \citep[e.g.][]{2007prpl.conf..245A}.

As the total scalar momentum, $P$, which each outflow should introduce is known, 
$v_j(R)$ can be expressed in terms of $P$ and the integral part of Eq.\,\ref{P}.
However, by setting a new velocity field, the total momentum of the outflow may not be conserved due to a 
non-uniform density distribution existing within the chosen outflow radius, especially during later times of the simulation.
Therefore we calculate the added momentum and then subtract it within the radius $R$.
This ensures the vector sum of the momentum is zero and the total added momentum on the computational domain remains conserved.

In order to give physical meaning to our model,
we make use of the same dimensionless parameter scheme as first introduced by \citet{2007ApJ...659.1394M}
and later used by \citet{2009ApJ...695.1376C,2010ApJ...722..145C}.
Following this scheme conveniently links our results to physical quantities and enables comparison of our results to those
of different codes.
We thus define in our model the dimensionless quantities of mass, $m_0$, length, $l_0$, and time, $t_0$, as,
\begin{equation}
 m_0 = \frac{\rho^{\frac{4}{7}} P^{\frac{3}{7}} } { S^{\frac{3}{7}} }, \mbox{ } l_0 = \frac{P^{\frac{1}{7}} } { \rho^{\frac{1}{7}}  S^{\frac{1}{7}} }, \mbox{ and } t_0 = \frac{\rho^{\frac{4}{7}}}{ P^{\frac{3}{7}}  S^{\frac{4}{7}} }.
\end{equation}
We assign the values of density $\rho$ = $2.51\times{10}^{-20}$ g\,cm$^{-3}$,
outflow momentum $P$ = $3.98\times{10}^{39}$ g\,cm$^{-3}$\,s$^{-1}$, and outflow rate
per unit volume $S$ = $6.31\times{10}^{-68}$ cm$^{-3}$\,s$^{-1}$, which were chosen
by \citet{2010ApJ...722..145C} to approximate the physical values in a typical star forming region.
After implementing the above unit scheme, the resulting physical unit quantities that we use in our code are;
$m_0$ = $18.7$ M$_{\odot}$, $l_0$ = $0.37$ pc, $t_0$ = $0.34$ Myr, and $v_0$ = ${l_0}$/${t_0}$ = $1.064$ km\,s$^{-1}$.
The local sound speed has a value of 0.587 km\,s$^{-1}$.
We set our domain scale length as 8\,$l_0$ (3.96 pc), 
choose an outflow radius to be $R$ = 0.3\,$l_0$, and run the simulation for time 3\,$t_0$\,($\sim$1.02 Myr). 

We perform the simulations on a fixed uniform grid of 512$^3$ cells with periodic boundary conditions 
to emulate a subset of a molecular cloud.
Numerical convergence of the density PDF data was tested at lower resolutions of 128$^3$ and 256$^3$. 
Although the majority of the cells provide precise convergence forming the peaks of the density PDF, 
there is a slight discrepancy in the tails. 
The same phenomenon was noted by \citet{2010A&A...512A..81F} who decides it is difficult to obtain 
exact numerical convergence through such turbulence simulations.  
However, we conclude a resolution of 512$^3$ is sufficient for our purposes as both the fitted mean and standard 
deviations of the PDF at all resolutions agree to four decimal places.

Finally, the model was implemented using an isothermal version of the {\sc tvd}
multi-dimensional fixed-grid code by \citet{1999ApJ...514..506K}.
The code is based on an explicit finite-difference scheme which uses a second-order accurate Roe-type up-winded
Riemann solver to calculate the inter-cell flux.
The performance of the code has been compared with several other prominent hydrodynamic codes in use
today in a study by \citet{2009A&A...508..541K}.
The study included three of the major grid-based codes ({\sc enzo}, {\sc flash} and {\sc zeus}) and compared simulations of
decaying, isothermal, supersonic turbulence using the same initial conditions.
In all cases, the {\sc tvd} code was found to give comparable results to its contemporaries but with substantially
shorter run-times.

\section{Results}
\label{results}

\begin{figure}
  \epsfxsize=8cm
    \epsfbox[45 25 500 260]{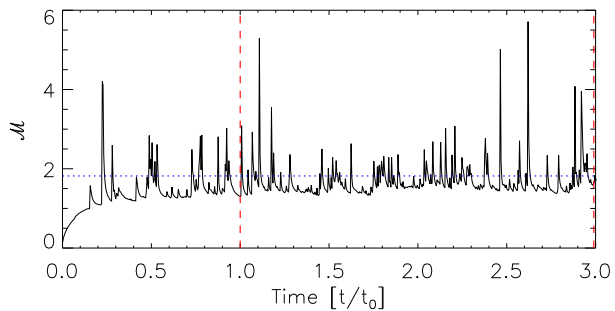}
\caption[]
{ Variation of the root mean square Mach number, {$\mathcal M$}, with time during our simulation.
The outflows quickly generate turbulent motions which saturate after a time of 0.5\,$t_{0}$.
The range over which we produce our time-averaged density PDFs is highlighted between the vertical 
dashed lines between time 1.0 to 3.0\,$t_{0}$.
The horizontal dotted line represents the Mach number corresponding to the outflow velocity scale, $v_0$.}
  \label{rmsplot}
 \end{figure}

The evolution of the root mean square Mach number, {$\mathcal M$}, 
during the 3.0\,$t_{0}$ running time is shown in Fig.\,\ref{rmsplot}.
The high velocity spikes visible in Fig.\,\ref{rmsplot} are a consequence of our method of implementing the outflows.
An outflow radius may include some cells of very low-density, and through the
conservation of momentum these cells would have very high-velocity.
We can control the magnitude of the high-velocity spikes by increasing the lower-threshold value 
where an outflow can be placed, but we compromise to ensure the outflows may be placed as randomly as
possible. The very high-velocity cells are localised and have little effect on the global turbulence evolution.
The outflows themselves are also quite localised so 
although normally possessing an initially high Mach number of ${\mathcal M}\sim20$, the global 
average {$\mathcal M$} is much lower, yet still supersonic at $\sim$1.7.

\begin{figure}
  \epsfxsize=8cm
    \epsfbox[1 50 540 550]{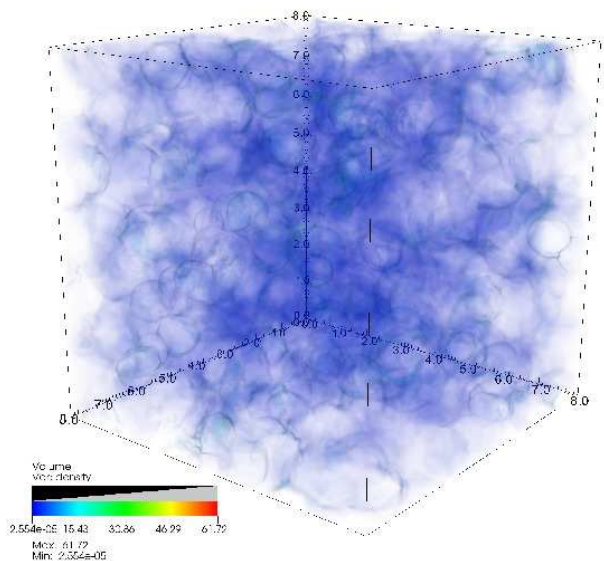}
\caption[]
{A 3D volume rendering of our simulation depicting density at time 2.0\,t$_{0}$. 
Despite a large density contrast from 2.55$\times$10$^{-5}$ to 61.72\,$\rho_0$, 
the majority of the volume is occupied by low-density spherical cavities interspersed by filamentary structures.}
\label{density_projection}
\end{figure}

A volume rendering of density during our simulation is shown 
at a time of 2.0\,$t_{0}$ in Fig.\,\ref{density_projection}.
It is apparent that the majority of the volume consists of gas at low-density with some spherical outflow cavities visible.
\begin{figure}{\centering}
   \epsfxsize=8cm
   \epsfbox[15 15 500 500]{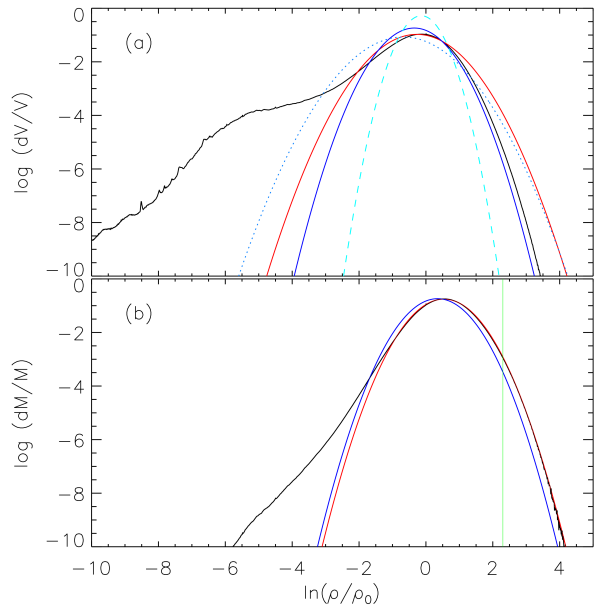}
\caption[]
{ {\bf (a)} Normalised density PDF plot in log.
Black line represents the time-averaged Volume fraction.
Red curve represents the best log-normal fit to the data using the Levenberg--Marquardt method.
Blue curve is a log-normal approximation assuming Fourier-based driving at the same Mach number.
Dotted dark-blue and dashed light-blue curves are the log-normal approximations of purely 
compressible and solenoidal driven turbulence respectively.
{\bf (b)} Same as (a) but for Mass fraction with a vertical green line to represent the threshold 
density for the CFR$_{ff}$ measurement.}
\label{pdf}
\end{figure}
To analyse the data quantitatively, we plot normalised Volume and Mass weighted density PDFs in natural-log.
Fig.\,\ref{pdf}a is the time-averaged Volume weighted density PDF (from 25 data dumps) 
between a simulation time of 1.0--3.0\,$t_{0}$.
Fig.\,\ref{pdf}b is the corresponding Mass weighted density PDF.
The solid black lines represent the density PDF of the normalised data arrays, $p_s \left( s \right)$, where 
$s$ = ln$\left( \rho / \rho_0 \right)$. 
We measure the Mean, $\mu$ = $\bar{s}$ = $\int s\,p_s \left( s \right) \mathrm{ds}$, 
and Standard Deviation, $\sigma^2$ = $\int {\left( s - \bar{s} \right)}^2 p_s \left( s \right) \mathrm{ds}$ of the distributions.
We also determine the best log-normal fit to the distributions using the Levenberg--Marquardt method. 
The resulting log-normal fits are plotted as the red curves. 
A log-normal distribution being defined as:
\begin{equation}
 f \left( s \right) = \frac{1}{\sqrt{2 \pi \sigma_{LN}^2}}exp \left[\frac{ - {\left( s - \mu_{LN} \right)}^2 }{2 \sigma_{LN}^2}\right]
\end{equation}
where $\mu_{LN}$ and $\sigma_{LN}^2$ are the Mean and Standard Deviations of the log-normal fit. 

Also determined are the values of Skewness and Kurtosis.
Skewness is a measure of symmetry of the distribution and defined as 
${\mathcal S}_s$ = $\int \left[ {\left( s - \bar{s} \right)/\sigma} \right]^3 p_s \left( s \right) \mathrm{ds} $.
Kurtosis measures the extent to which a distribution deviates from a normal-distribution and is defined as
${\mathcal K}_s$ = $\int \left[ {\left( s - \bar{s} \right)/\sigma} \right]^4 p_s \left( s \right) \mathrm{ds} -3$, 
where the last term is often included to ensure the Kurtosis of a standard normal distribution is 0.
The measured quantities are listed in Table \ref{model_parameters}. 

The negative skewness values indicate the density PDFs are skewed to the low-density side, 
although the skewness of the Mass weighted density PDF is closer to log-normal. 
This could be expected due to the mass of the diffuse material having less effect on the low-density side of the density PDF. 
The Kurtosis measurements show the same trend while being positive 
or `leptokurtic' which indicate an enhancement in the density PDF tails.

Comparing the density PDFs to their log-normal fits in Fig.\,\ref{pdf}, 
we see a distinctive low-density tail on the low-density or diffuse side, pronounced by the log-based plot.
It is due to the spherical outflows of our model which have expanded outwards leaving lower density `fossil cavities'.
In our simulations we see how fossil cavities do enhance the transfer of momentum to the ambient medium. 
Through the conservation of momentum, material moves faster in the low-density regions, thus 
enabling the outflow shocks to propagate further from the source through older cavities than they 
would otherwise travel in an undisturbed uniform medium.

Due to the fact we do not implement self-gravity, we do not obtain an enhanced
high-density tail on the PDF as found by other authours \citep[e.g.][]{2011MNRAS.410L...8C, 2013ApJ...763...51F}.
Although we find the Mass weighted density PDF is a better match to the high-density side of the log-normal fit than the 
Volume weighted equivalent. Again, this is due to the Volume weighted density PDF being skewed towards lower density by a 
greater amount.

We also plot a log-normal curve to represent the result one would obtain from Fourier driven turbulence at 
the same average Mach number as our simulation.
We proceed by making use of the forcing parameter, $\zeta$, originally employed by \citet{2008ApJ...688L..79F} as the relative strength of the 
solenoidal to compressional components. 
Although the authors defined it with respect to their forcing function, we define it at the end of our simulation 
based on the resulting turbulent velocity field averaged over the time interval 1.0--3.0\,$t_0$. 
We measure $\zeta$ to be 0.614, a value that indicates solenoidal modes slightly dominate in our 
fully converged saturated turbulent velocity field.
Then through equation 5 of \citet{2008ApJ...688L..79F} we determine the corresponding $b$ value (0.591) and use it, 
along with a measure of our average Mach number (1.69), in the $\sigma^2 = \mathrm{ln} \left(1+b^2 \mathcal{M}^2\right)$ equation, 
and the $\mu = -\sigma^2/2$ relation of \citet{1998PhRvE..58.4501P}, 
to obtain a standard-deviation and mean value for plotting a general Fourier-driven based log-normal approximation.
In Fig.\,\ref{pdf}a we also plot the log-normal curves corresponding to the two extremes; 
pure solenoidal driving ($\zeta$=1.0) and pure compressible driving ($\zeta$=0.0).

From visual inspection, we see the Fourier-driven based log-normal approximation is a more precise fit to
the high-density end of the Volume weighted density PDF than its best log-normal fit. 
In the Mass weighted density PDF, the Fourier-driven based log-normal approximation is a closer match to the best log-normal fit
(see standard deviation values in Table\,\ref{model_parameters}).

The departure of our data from the standard log-normal shaped density PDF is most likely due to the fact
that we drive the turbulence in real-space using outflows, whereas most numerical works who 
study density PDFs drive the turbulence in Fourier-space based on Gaussian distributions 
\citep[e.g.][]{2008ApJ...688L..79F, 2012ApJ...761..156F, 2012MNRAS.423.2680M}.
By driving turbulence in Fourier-space, the material over the entire domain is mixed in a sinusoidal based manner; 
some portion becomes high-density, and an equal portion becomes low-density. 
It is hard to obtain a significantly {\it non} log-normal density PDF in this way 
without extra physics such as self-gravity.
We would expect a difference by driving the turbulence in real-space, as our spherical outflow model is highly 
localised. 
The outflows are introduced to regions of above average density and easily inflate 
low-density bubbles in the computational domain.

The relationship between the star formation rate and density PDF has been investigated by several 
groups, namely, \citet{2005ApJ...630..250K}, \citet{2011ApJ...741L..22P} and \citet{2011ApJ...743L..29H}. 
Each group used a slightly different theoretical measure to define the critical density for star formation to occur 
and thus estimates different values for the star formation rate.

Here we investigate the theory of CFR$_{ff}$.
In our Mass weighted density PDF plot of Fig.\,\ref{pdf}b we choose a critical density of ${\rho}$/${\rho_0}$ = 10 
and calculate the fraction of the PDF area greater than this density.
This is equivalent of implementing the Error Function as defined by 
equation 20 of \citet{2005ApJ...630..250K} with $\epsilon_{core}$ = 1, and later used by  \citet{2011MNRAS.410L...8C}.
Although our model is not entirely appropriate for a precise estimate of the CFR$_{ff}$ 
as we do not implement self-gravity and so can not take a critical density from the 
Jeans Length measurements as in \citet{2005ApJ...630..250K},
here we simply take a critical density independent of Mach number and
can still make a comparison of this arbitrary CFR$_{ff}$ between the Mass weighted density PDF and its 
corresponding log-normal fittings.

The CFR$_{ff}$ values are listed in Table\,\ref{model_parameters}. 
The results show that CFR$_{ff}$ of the Fourier driven log-normal fit based on the $\zeta$ value determined from our 
resulting velocity fields, differs from that of the outflow driven data by a factor of $\sim$2.
But as we are using a slightly different definition of $\zeta$ as explained above, 
a factor of 2 difference may still be in good agreement.
Overall, a log-normal approximation to the data, or a Fourier-driven turbulence model could still 
provide comparable CFR$_{ff}$ results with our particular real-space driving model as the CFR$_{ff}$ is dependent
on mass measurements at high-densities.
We will be able to perform a more accurate analysis in a future publication with self-gravity implemented.

\begin{table}
\centering
\caption{Summary of measurements obtained from the density PDFs of the outflow driven (OD) data, the best log-normal (LN) fits, 
and the Fourier-driven based log-normal approximation (FD), for both the Volume and Mass weighted cases. 
Bottom row shows the estimated values of the CFR$_{ff}$, which is a measurement based on mass.}
\label{model_parameters}
\tabcolsep=0.15cm
\begin{tabular}{@{} l c c c c c c}
\hline
                &  \multicolumn{3}{c}{(a) Volume weighted} & \multicolumn{3}{c}{(b) Mass weighted} \\
                           &  OD   & LN      & FD       &  OD   & LN    & FD    \\
\hline
Mean $\mu$                 & -0.732 & -0.272  & -0.348   & 0.501  & 0.539 & 0.348 \\
Std-dev $\sigma$           &  1.560 &  1.057  &  0.834   & 0.896  & 0.845 & 0.834 \\
Skewness ${\mathcal S}_s$  & -1.490 &  0.0    &  0.0     &-0.326  & 0.0   & 0.0   \\
Kurtosis ${\mathcal K}_k$  &  3.016 &  0.0    &  0.0     & 0.99   & 0.0   & 0.0   \\
CFR$_{ff}$                 &        &         &          & 0.0185   & 0.0189  & 0.0097\\ 
\hline
\end{tabular}
\end{table}
\section{Conclusions}
\label{conclusion}
We present simulations of outflow driven turbulence performed using the isothermal {\sc tvd} code. 
The turbulence is driven exclusively in real-space by a series of spherical outflows 
based on the parameter scheme as used by \citet{2007ApJ...659.1394M} and \citet{2009ApJ...695.1376C,2010ApJ...722..145C} 
in order to approximate the physical values of a typical of star-formation cloud. 

We analyse the resulting turbulence in terms of Volume and Mass weighted density PDFs, and the CFR$_{ff}$
over a driven regime in which outflows are continuously introduced.
Our primary finding is a departure from log-normality of the density PDF due to an extended low-density tail, 
quantified by measures of negative Skewness and positive Kurtosis.
It is a consequence of our localised real-space driving method where the multiple outflows sweep the ambient gas into thin shells, 
thus enhancing the volume of low-density cavities.
We propose that if there is localised turbulent driving in an 
active cloud via an outflow, stellar wind, or supernova, then the density PDF may have an enhanced tail at the low-density side.
However, log-normal fits to the data may still provide good approximations to the high-density side 
especially when considering Mass weighted density PDFs rather than Volume weighted.

Although we do not implement self-gravity, we quantify our results with a CFR$_{ff}$ calculation which 
we define as the fraction of density PDF area greater than ${\rho}/{\rho_0} = 10$, 
as similarly implemented by \citet{2011MNRAS.410L...8C}.
Our results suggest that the CFR$_{ff}$ based on our particular simple real-space driving model with $\zeta\sim0.6$ 
is comparable to the CFR$_{ff}$ that would be obtained through Fourier-driving at the same Mach number.

In future work, we shall perform a power-spectra analysis of higher-resolution simulations and 
a more detailed analysis of the compressional and solenoidal generated by our real-space driving mechanism. 
We will also improve our physical model by implementing self-gravity so that dense cores form naturally, add 
collimation to the outflows, and include the effects of magnetic fields as are present in molecular clouds.

\section*{Acknowledgements}
Numerical simulations were performed using a high performance computing
cluster in the Korea Astronomy and Space Science Institute (KASI).
This work is partially supported by the KASI-Yonsei Joint
Research Programme (2011-2012) for the Frontiers of Astronomy and
Space Science funded by the KASI, 
and the DRC program of Korea Research Council of
Fundamental Science and Technology(FY 2012).
AM thanks Young-Min Seo for assistance and useful discussions.
S.-J.Y. acknowledges support
from Mid-career Research Program (No. 2012R1A2A2A01043870)
through the NRF of Korea grant funded by the MEST,
by the NRF of Korea to the Center for Galaxy Evolution Research (No.
2010-0027910), and by the KASI Research Fund 2012 and 2013.
The authours thank the referee for useful comments and suggestions.

\footnotesize{
\bibliographystyle{mn2e}
\bibliography{tvd}
}

\label{lastpage}

\end{document}